\documentclass[12pt,a4paper]{article}
\usepackage[english]{babel}
\usepackage[latin1]{inputenc}
\usepackage[T1]{fontenc}
\usepackage{times}

\usepackage{amsthm,amsfonts,amsmath}
\usepackage{graphicx,graphics}
\usepackage{bbm,url}

\newcommand{\Z}{{\mathbb Z}}
\newcommand{\R}{{\mathbb R}}

\newcommand{\vc}[1]{{#1}} 
\newcommand{\vep}{\varepsilon}

\newcommand{\defem}[1]{{\em #1\/}} 
\newcommand{\mean}[1]{\langle #1\rangle}
\newcommand{\selfc}{self-consistent}
\newcommand{\ie}{{\em i.e.\/}}
\newcommand{\etc}{{\em etc\/}}
\newcommand{\eg}{{\em e.g.\/}}
\newcommand{\coleq}[1]{{#1}} 
\newcommand{\jecurr}{j^{\rm e}}
\newcommand{\jrcurr}{j^{\rm r}}
\newcommand{\mcss}[1]{\overline{\mathcal{#1}}}
\newcommand{\rme}{{\rm e}}
\newcommand{\rmd}{{\rm d}}

\newcommand{\email}[1]{E-mail: \tt #1}
\newcommand{\emailjani}{\email{jani.lukkarinen@helsinki.fi}}
\newcommand{\addressjani}{\em University of Helsinki,
Department of Mathematics and Statistics\\
\em P.O. Box 68,
FI-00014 Helsingin yliopisto, Finland}

\begin{document}

\title{Energy fluctuations, hydrodynamics and local correlations in harmonic systems with bulk noises}

\author{Jani Lukkarinen\thanks{\emailjani}\\[1em]
$^*$\addressjani}

\maketitle

\begin{abstract}
In this note, I summarise and comment on joint work with C.~Bernardin, V.~Kannan, and J.~L. Lebowitz concerning two harmonic systems with bulk noises whose nonequilibrium steady states (NESS) are nearly identical (they share the same thermal conductivity and two-point function), but whose hydrodynamic properties (convergence towards the NESS) are very different.  The goal is to discuss the results in the general context of nonequilibrium properties of dynamical systems, in particular, what they tell us about possible effective models, or \defem{predictive approximations}, for such systems.
\end{abstract}

%\pacs{02.50.-r, 05.20.Dd, 05.40.Ca, 05.50.+q, 05.60.Cd, 05.70.Ln}

\maketitle

\section{Introduction}

This note is mainly based on the results derived in \cite{bkll11b} and summarised from a more physical point of view in \cite{bkll11}.  Both are made in collaboration with C.~Bernardin, V.~Kannan, and J.~L.~Lebowitz; hence the discussion here relies heavily on the work of others, although naturally the responsibility for any possible faults is solely mine.  

Two dynamical systems are considered in these works, combining the same harmonic Hamiltonian evolution with two different types of bulk noise.  
We consider $d$-dimensional crystals with the harmonic evolution determined by nearest neighbour interactions on a square lattice with fixed boundary conditions in the first direction and periodic boundary conditions in all other directions (if $d\ge 2$).  The geometry is chosen mainly for computational convenience. It has two natural boundaries, called the two \defem{ends} of the crystal, formed out of the sites attached to the fixed boundary particles.  We denote the number of particles in the first direction by $L$ and use the lattice spacing as the microscopic length unit: for instance, the microscopic volume is equal to the number of particles, $N=L^d$.  By a \defem{thermodynamic limit} of such systems, I mean taking $L\to \infty$.

Both of the above models have been studied before: the first one uses the \defem{\selfc\ heat baths} as in \cite{rich75,bll02}
while the second \defem{velocity flip model} has been considered in \cite{FFL94,BO11,dkl11}.  In the \selfc\ model, each particle is coupled to a Langevin heat bath and the
input temperatures of the baths can vary along the chain but are uniquely determined by fixing the temperatures on the left and right end of the system at some given values, $T_{\rm L}$ and $T_{\rm R}$, and then requiring that in the remaining sites there is no energy flux \defem{on average} between a particle and its heat bath \defem{in the stationary state}.  As shown in \cite{bll02},
this requirement leads to a system with a unique nonequilibrium steady state (NESS) satisfying the stationary Fourier's law.  We also obtain explicit formulae for the heat conductivity and for the dependence of correlations on the \selfc\ temperature profile.

The bulk noise in the \selfc\ model only conserves energy on average, and even then only in the NESS.  
The second bulk noise acts less disruptively, by randomly flipping the signs of particle velocities, and thus conserving energy in each realisation of the noise.
More precisely, one can imagine that each particle carries its own (Poissonian) clock whose rings will enforce a velocity reversal, independently of what the other particles are doing.  We then attach Langevin heat baths at the ends of the crystal, just as in the \selfc\ system above.  It is shown in \cite{dkl11} that this model converges to a unique NESS which has a covariance matrix and thermal conductivity \defem{identical} to that of the \selfc\  model, as long as the parameters of the harmonic evolution and the boundary conditions match.

Therefore, it might sound reasonable to conjecture that the thermal conduction properties of the two systems are identical, even though both noises are ``strong perturbations''  in the sense that they convert the pure harmonic system with infinite conductivity \cite{lebo67} into a normally conducting one.  In fact, our results strongly indicate that the NESS:s of these two systems are \defem{locally} indistinguishable from each other: the difference between the NESS expectations of any local observable vanishes in the thermodynamic limit.  However, they are not \defem{globally} identical since, for instance, the total energy fluctuations in the NESS differ by an amount which diverges as $L\to \infty$.  This turns out to be due to long-range corrections to local thermal equilibrium expectation values which are found in the velocity flip model but which are absent from the \selfc\ model.  
Even more pronounced is the difference in the convergence towards the NESS: for the \selfc\ model with pinning this occurs microscopically fast (exponential decay which is $O(1)$ in the original microscopic time scale) whereas it will typically take $O(N^2)$ microscopic time units before the state of the velocity flip model is close to its NESS.  

The purpose of this note is to recall the arguments why the latter behaviour should be the one most commonly found in physical systems, and to discuss how our earlier results can be interpreted as \defem{effective models} for the energy transport in these systems.
Its outline follows closely a talk given by the author at a workshop in the NORDITA program \defem{Foundations and Applications of Non-equilibrium Statistical Mechanics} in 2011. Here I will try to clarify the key concepts and terminology in two explanatory sections, \ref{sec:lte} and \ref{sec:scalings}, preceding the related results.

\section{On local thermal equilibrium}\label{sec:lte}

The intuitive meaning of the terms \defem{local} and \defem{global} is clear: ``local'' refers to the properties of the system in a (microscopic) neighbourhood of a point and ``global'' concerns the properties of the entire system.  The distinction becomes relevant only for large systems, when the entire system spans a large spatial region and these two length scales thus separate; in other words, in the thermodynamic limit.

A local equilibrium state should thus be a state in which the properties of the system near a given point $x$ in space are determined by some equilibrium state. 
For the present purpose, I will only consider \defem{equilibrium states} determined by ``canonical Gibbs measures'': given a collection of $n$ conserved quantities $O_i(q,p)$, $i=1,\ldots,n$, $q,p\in (\R^{d})^N$, the corresponding canonical Gibbs measures are labelled by $n$ real numbers $\lambda_i$ such that the measure $\rme^{\sum_i \lambda_i O_i(q,p)} \rmd q \rmd p$ is normalisable.  As long as the Lebesgue measure $\rmd q \rmd p$ is invariant under the time-evolution---which is the case for any Hamiltonian evolution, also together with the above velocity flips---any of such Gibbs measures is stationary.  \defem{Thermal equilibrium states} are then defined by choosing as the observables the total energy, $O_1(q,p):=H(q,p)$, and any other thermodynamically relevant conserved quantities, such as particle number, total momentum, \etc . \defem{Temperature} is then defined by $T:=\beta^{-1}$ where $\beta:= -\lambda_1$ typically needs to be positive for the measure to be normalisable.  These states also serve as good reference measures of the \selfc\ heat bath model in the sense that if all baths have the same temperature then the corresponding Gibbs measure is invariant under the time-evolution.

We say that a state $\mu$ of the system is in local $\mu_0$-equilibrium at $x$, if $\mu_0$ is an equilibrium state of the (infinite) system and $\mean{F(q,p)}_\mu\approx \mean{F(q,p)}_{\mu_0}$ for all observables $F$ localised at $x$.  More precisely, this should hold for any $F$ which depends only on $q_j,p_j$ for those $j$ with $|x-x_j|<R$ where $R>0$ is a large microscopic length and $x_j$ is the spatial location of particle $j$ (usually, $x_j=q_j$ but for our crystals we use $x_j=j$).  The correction term should vanish in the thermodynamic limit, and $R$ can then be arbitrary as long as it is $L$-independent.  The state $\mu$ is a \defem{local equilibrium state} if this property holds at every point $x\in \R^d$.  

Thus to have a local equilibrium state necessarily implies the existence of a collection $\mu_0(x)$ of equilibrium states labelled by points $x$, although in the present generality, $\mu_0(x)$ are not uniquely determined.  If the reference equilibrium states are chosen from the thermal equilibrium states, then the state is said to be in \defem{local thermal equilibrium} (LTE) and the distribution of the parameters $\lambda_i(x)$ is typically unique up to errors which vanish in the thermodynamic limit.  In particular, if we allow some thermodynamically small errors to the \defem{temperature distribution} $T(x)$, its meaning is then uniquely determined for any LTE system.  Note that although then always $T(j)=\mean{p_j^2}$ for any LTE state of a standard Hamiltonian system for which energy is the only conserved quantity, the above definition implies more: then necessarily also $T(j+r)\approx T(j)$ for any microscopic lattice displacement $r$.  Thus LTE implies that $T(x)$ must be essentially constant in any microscopic region of such systems.

LTE states are common in simulations and in strongly stochastic particle systems, and it even seems reasonable to conjecture that in great generality an initial state will converge into some LTE state in a finite microscopic time.  Indeed, it is questionable if the term ``temperature distribution'' should ever be applied to a state which is not at LTE.
However, rigorous proofs of such generic convergence towards LTE have only been achieved in certain special cases, and it could well be the hardest part of a mathematical proof of dynamical Fourier's law.  For instance, we have a proof of LTE in the present \selfc\ model, which has a \defem{harmonic} Hamiltonian term, but it remains the only part missing from an analogous proof when the Hamiltonian is anharmonic \cite{bllo09}.

\section{Local and global correlations: how can something be both vanishing and essential?}

How do the correlations behave in the two models described in the introduction?  For simplicity, I will only consider the one-dimensional case in detail; some of these results immediately generalise to higher dimensions, as explained in \cite{bkll11b}.  Thus, from now on, $L=N$ and $d=1$.

The Hamiltonian part of the dynamics is harmonic, and thus by itself it has order $N$ conserved quantities and it transports energy ballistically through the system.  In particular, it has an infinite thermal conductivity.  Our two bulk noises have been chosen precisely so that they would break most of these conservation laws without breaking the conservation of the total (Hamiltonian) energy of the system.  In the \selfc\  case the energy is conserved only on average in the NESS (this is achieved by the \selfc\  tuning of the bulk heat bath temperatures) and in the velocity flip model it is conserved in the bulk with probability one.  However, as will be seen shortly, also other relevant conserved quantities can appear.

As mentioned in the introduction, both systems have a unique NESS, and the two-point functions of these states, \ie , the covariance matrices of the variables $\{q_j,p_j\}$, coincide.  The NESS of the \selfc\  model is Gaussian, and thus uniquely determined by the covariance matrix.  The NESS of the velocity flip model has some nonzero fourth order cumulants and thus cannot be Gaussian.  Nevertheless, our results also indicate that these higher order cumulants all approach zero as $N\to\infty$.  If this is true, all local correlations would agree in the thermodynamic limit, and 
it follows from the known properties of the \selfc\ case that both NESS:s must then be LTE states with the \defem{same} temperature distribution. 

However, the convergence towards equilibrium is radically different in these two models.  
In the \selfc\  case with pinning, whatever the initial state, all local expectations relax to their NESS value  exponentially fast on a microscopic time-scale, while in the velocity flip case local energy needs a time $O(N^2)$ to equilibrate.  In addition, even though they approach zero in the thermodynamic limit, the above mentioned corrections to LTE do contribute towards energy fluctuations in the NESS (this can happen since the energy fluctuations are determined by a sum over $O(N^2)$ local correlation functions).  The next subsection summarises our findings.

\subsection{Comparison of energy fluctuations at the NESS} \label{sec:energyfluc}

In an infinite system without boundary terms,
the bulk dynamics of the velocity flip model conserves the energy density, $H(p,q;N)/N$, $N\to\infty$.  
The Hamiltonian function of the finite chain is given by
\[
H(\vc{p},\vc{q}) = \sum_{j=0}^{N+1} \mathcal{E}_j =
\smash{\sum_{j=0}^{N+1}} \Bigl( \frac{p_j^2}{2}  + \nu^2 \frac{ q_j^2}{2} +
 \sum_{i:|i-j|=1}\frac{(q_{j}-q_i)^2}{4} \Bigr) \, ,
\]
where $\mathcal{E}_j=\mathcal{E}_j(\vc{p},\vc{q})$ denote the \defem{local energy observables}.
Here we say that the system has \defem{pinning} whenever $\nu\ne 0$.  Without pinning ($\nu=0$), also the \defem{total deformation density}, $N^{-1}\sum_{j} r_j$, $r_j := q_{j+1}-q_{j}$, is conserved and can fluctuate.  (Since it is always, also with pinning, equal to $q_{N+1}-q_0$, it is trivially conserved in the finite system by the chosen boundary conditions.  However, the point is whether its value can vary in the infinite volume canonical Gibbs state: this is possible if $\nu=0$ but not if $\nu\ne 0$.)

In contrast, the bulk dynamics of the \selfc\ model does not preserve the energy density, although without pinning it does preserve the deformation density.
This model is explicit enough to be studied rigorously, even in the thermodynamic limit: I quote below some of the results proven in \cite{bkll11b} (without pinning) and in \cite{bll02} (with pinning).  

The correlations at NESS depend linearly on the input temperatures of the heat baths, 
$\mean{X_i Y_j} = \sum_{n=1}^N B_{XY}^{(n)}(i,j) T_n$, $X,Y\in \{q,p\}$.  Here $B$ decays for increasing $M:=1+|i-j|+|i-n|+|j-n|$, at least as fast as dictated by the following bounds:
\begin{description}
 \setlength{\itemsep}{0pt}
 \item[With pinning,] $|B_{\cdot\cdot}^{(n)}(i,j)|\le c \rme^{-a M}$ for some $c,a>0$ independent of $N$.
 \item[Without pinning,] for some $N$-independent $c>0$ and denoting $r_j=q_{j+1}-q_j$,
\begin{align*}
& | B_{rr}^{(n)}(i,j)|\le c M^{-2} (1+\ln M)\, , \quad
| B_{rp}^{(n)}(i,j)|\le c M^{-3} (1+\ln M)\, , \\ &
| B_{pp}^{(n)}(i,j)|\le c M^{-4} \, .
\end{align*}
\end{description}
These results also yield the following estimate on energy fluctuations in the \selfc\  NESS,
\[
   \frac{1}{N} \mean{H;H} =   \frac{1}{N}\sum_{j,j'=0}^{N+1} \mean{\mathcal{E}_{j'}; \mathcal{E}_{j}}^{({\rm eql},T_{j})} + R_N \, ,
\]
where we denote $\mean{A;B} = \mathrm{Cov}(A,B) = \mean{A B}-\mean{A}\mean{B}$ and 
\begin{description}
 \setlength{\itemsep}{0pt}
 \item[with pinning] $R_N=O(N^{-\frac{1}{2}})$,
 \item[without pinning] $R_N=O(N^{-\frac{1}{4}} \ln^2 N)$.
\end{description}
(The bounds for $R_N$ are likely off by a power $N^{-\frac{1}{2}}$ due to our poor control over the dependence of the \selfc\  temperature profile on $N$.)
Thus we can conclude that the energy fluctuations in the \selfc\ model are $O(N)$ and only the LTE expectations contribute to the leading behaviour.

At the moment, we do not have any mathematically rigorous bounds for the energy fluctuations at the NESS of the velocity flip model, but numerical simulations give clear evidence that, although of same magnitude in $N$, they do not agree with the above LTE result.  For instance, if $T_{\rm L}=1$ and $T_{\rm R}=8$, the LTE contribution to the $N\to\infty$ limit of $s_N:=N \mean{H;H}/\mean{H}^2$ can be rounded to $1.20$, while simulations produce $1.40(2)$ consistently for both $N=200$ and $N=400$  (more results are given in \cite{bkll11}).  A possible mechanism, yielding an approximate value $1.40$ in this case, is explained in Section \ref{sec:flucthydro}.

\section{On kinetic and hydrodynamic scaling limits}
\label{sec:scalings}

Various scaling limits have turned out to be useful in the study of conduction properties of dynamical systems.   They share the feature that by considering a restricted collection of observables
and an ``unphysical'' limit for some of the parameters of the dynamics, one obtains more easily solvable evolution equations for the limit observables.  Therefore, such a limit 
produces an ``effective theory'' 
whose appropriately rescaled solutions yield approximations to the chosen observables of the original system; this typically under circumstances where a direct solution of the original system is not possible.  

The term ``effective theory'' is commonly used in physics but, in my experience, it takes some effort to explain the concept to those unfamiliar with it. After all, it is ``effective'', not because it is very accurate (usually rather the opposite) but because it bypasses some obstacle, and it is not really a theory at all but a model for some parts of the original ``theory''.
Personally, I would prefer to replace this term with something more immediately descriptive, such as \defem{predictive approximation}: the result is an approximation which can be used to predict some otherwise uncontrollable properties of the original system.

Examples of such scaling limits abound: mean field limits, adiabatic limits, renormalisation group transformations, \etc .  Here I will only discuss two such limits in detail: the hydrodynamic and kinetic scaling limits.  In a \defem{hydrodynamic scaling limit} space is scaled by $L^{-1}$ and time by $L^{-2}$.  For nonequilibrium problems $L$ is usually chosen so that the size of the system remains finite in the limit, and thus it can have boundaries through which the nonequilibrium state can be generated.  For instance, a solution $o(x',t')$ of a predictive approximation could be related to a solution of the original system by
\begin{align}
& \int_{\R^d} \rmd x' \int_0^\infty \rmd t'\, F(x',t') o(x',t') \nonumber\\ & \quad \approx L^{-d} \sum_{j\in \Z^d, |j|<L/2} L^{-2} \int_0^\infty \rmd t\, F(L^{-1} x_j(t),L^{-2} t) \mean{O_j(t)}\, ,
\end{align}
for all large enough $L$. Here the observable $O_j(t):=O_j(q(t),p(t))$ measures some microscopic property carried by the particle $j$, $F$ is an arbitrary compactly supported test-function and $x_j(t):= x_j(q(t),p(t))$ denotes the spatial position of the particle $j$ at time $t$.  The hydrodynamic scaling limits are best suited to study diffusive phenomena, as the scaling leaves such invariant, and the resulting predictive approximations typically (and also here) involve diffusion processes.

The hydrodynamic scaling limits I would like to contrast with \defem{kinetic scaling limits}.  Although the latter term is not as firmly established as the first one, it can be motived by two properties shared by the limit evolution equations: First, the scaling leaves velocities invariant and the evolution will typically be dominated by constant velocity, \ie , ballistic motion.  Secondly, the resulting limit equations are often those found in kinetic theory, such as Boltzmann transport equations.

A common way to arrive at a kinetic scaling limit is to start with a system which has some explicitly solvable ``free'' dynamics involving motion with constant velocity.  Free classical particles form obviously such a system but so do many wave-equations, at least if one considers the evolution of the Wigner functions of the wave fields.  Then a perturbation is introduced into the system, for instance, by adding a potential term $\lambda V$.  In any case, let $\lambda>0$ denote the ``strength'' of the perturbation with $\lambda=0$ corresponding to pure free evolution. The scaling limit is then defined by scaling space and time with $\lambda^{a}$, $a>0$, and considering $\lambda\to 0$; this obviously leaves all velocities invariant.  Here different choices of $a>0$ typically produce different types of limit evolution.  However, most often there is a unique choice $a_0$ such that the limit equations are ballistic for any $a$ with $0<a<a_0$ but become nonballistic for $a=a_0$.
For instance, for many weakly perturbed wave equations $a_0=2$ (for rigorous results, see \eg\  \cite{erdyau99,ls05,NLS09}).

If the limit evolution is not diffusive, it is not clear how to use hydrodynamic scaling limits.
At the very least one needs to modify the scaling functions, but how?  Here kinetic scaling limits can be very helpful by identifying the first nontrivial effects produced by the perturbation (for any time scale $O(\lambda^{-a})$, $a<a_0$, the motion is ballistic, hence trivial).  Then one can study what happens to the predictive approximations---produced by rescaling the solutions of the kinetic limit equation---at times longer than the kinetic time scale.  If no new effects appear, the overall result can be a simple correction to the constants predicted by the kinetic equation.  This seems to be common for normally conducting systems where the diffusion constants computed from the hydrodynamic scaling limit of the kinetic equation determine the leading behaviour of conductivity as $\lambda\to 0$. However, the terms neglected in the kinetic scaling limit can also alter the character of evolution entirely.  For instance, the Boltzmann equation arising in the kinetic limit of the two-dimensional Anderson model is expected to be diffusive \cite{erdyau04} while the original model should be exponentially localised, with the localisation length diverging for $\lambda\to 0$ (for related experimental evidence of such phenomena, see \cite{and2d07}).

Thus one should not treat the ``effective theories'' obtained from the scaling limits as totally universal, but rather as predictive approximations which might persist for times longer than those indicated by the scaling, but do not necessarily do so.  One has to be particularly careful if the predictive approximation generates singularities.  These singularities might conspire with some of the neglected terms and affect the evolution already during finite kinetic time scales.  This appears to happen in Bose condensation in a bosonic Boltzmann-Nordheim equation \cite{spohn08}, and the ``entropy solutions'' to Euler equation provide another example of a nontrivial continuation beyond a singularity.

\section{Fluctuating hydrodynamics of the velocity flip model}
\label{sec:flucthydro}

Only the velocity flip model without pinning will be discussed here since its hydrodynamic structure is richer than the one with pinning or in the \selfc\ model.
The local quantities building up the energy and deformation densities, as well as the associated currents, are then
\begin{align*}
 & \mathcal{E}_j  = \frac{p_j^2}{2} +\frac{r_j^2}{4} + \frac{r_{j-1}^2}{4},\qquad & & \jecurr_j = -\frac{1}{2} r_{j-1} (p_j + p_{j-1})\, , \\ &
 r_j  =q_{j+1}-q_j,  \qquad && \jrcurr_j  = -p_j\, .
\end{align*}
Explicitly, the observables and their currents satisfy
$\mathcal{E}_j (t) - \mathcal{E}_j (0) = \int_0^{t}\! \rmd s  \left[ \jecurr_j (s) -\jecurr_{j+1} (s)\right]$ and $r_j (t) - r_j (0) = \int_0^{t}\! \rmd s  \left[ \jrcurr_j (s) -\jrcurr_{j+1} (s)\right]$.

We parameterise the canonical Gibbs measures in terms of $\beta=T^{-1}>0$ and $\tau\in \R$,
\begin{align*}
 \coleq{Z(\beta,\tau)^{-1} \prod_{j=0}^{N+1} \exp\Bigl[-\beta (\mathcal{E}_j - \tau r_j) \Bigr]}\, ,
\end{align*}
and denote expectations over this measure by $\langle \cdot \rangle_{T,\tau}$.  Then
\begin{align}
\langle p_j^{2} \rangle_{T,\tau} =T, \quad \langle \mathcal{E}_j \rangle_{T,\tau} = T + \tau^2 /2, \quad \langle r_j \rangle_{T,\tau} =\tau \, .
\end{align}

We next \defem{assume} that LTE with respect to the above measures holds for all times.  Then 
the \defem{mean} energy density $\vep(x',t')$ and deformation density $u(x',t')$, under hydrodynamic scaling, satisfy $\vep(x',t') = T(N x',N^2 t') + \tau(N x',N^2 t')^2 /2$ and $u(x',t')=\tau(N x',N^2 t')$, up to small errors.
For this particular model, there is also a happy accident, and the evolution equations of the hydrodynamic fields can be closed merely by assuming LTE:  Since ($\mathcal{L}$ denotes the generator of the stochastic process)
\begin{align*}
&\jecurr_j = - (\nabla \phi)_j + \mathcal{L} (h_j), \quad \jrcurr_j = -\gamma^{-1} (\nabla r)_{j-1} + \mathcal{L} (\gamma^{-1} p_j), \\
&\phi_j= (2 \gamma)^{-1} (p_{j-1}^2 +r_{j-1} r_{j-2}), \quad h_j = -\gamma^{-1} \jecurr_j \, ,
\end{align*}
the slightly formal argument in \cite{bkll11b} yields that  for $t>0$, $x\in [0,1]$ the fields $u(x,t)$ and $\vep (x,t)$ should satisfy
\begin{align}
%\begin{cases}
&\partial_t u = \gamma^{-1}\,  \partial_x^2 \, u\, , \label{eq:ueq}\\
&\partial_t {\varepsilon}  = (2 \gamma)^{-1} \, \partial_x^2 \, ( \varepsilon + u^2 /2) \, , \label{eq:eeq}
%\end{cases}
\end{align}
with the boundary conditions
\begin{align}
& (\partial_x u)(0,t) = (\partial_x u)(1,t)=0,\\
& \bigl(\vep -{u^2}/{2}\bigr) (0,t) = T_{{\rm L}}, \quad \bigl(\vep -{u^2}/{2}\bigr) (1,t)= T_{{\rm R}}\, . \label{eq:ebc}
\end{align}
In fact, in a recent preprint \cite{Simon12} an analogous result has been rigorously derived for slightly different boundary conditions and assuming that the initial state is close, in the sense of relative entropy, to an LTE state.
 
Thus the above result already provides a predictive approximation for the diffusive scale (\ie , macroscopic) evolution of mean energy and deformation densities in this system.  However, there are natural questions for which it is not directly predictive, the fluctuations of the total energy being one of them.  To study the fluctuations of the hydrodynamic fields around their expectation values $\vep$ and $u$, we define, for an arbitrary choice of test functions $F$ and $G$, the \defem{fluctuation field observables} as
\begin{align*}
\mathcal{R}_t^N(F) = \frac{1}{\sqrt N} \sum_{j=1}^N F(j/N) \left[ r_j (t N^2) - u(j/N, t) \right]\, ,\\
\mathcal{Y}_{t}^N (G) = \frac{1}{\sqrt N} \sum_{j=1}^N G(j/N) \left[ \mathcal{E}_j (tN^2) - {\varepsilon}(j/N, t) \right]\, .
\end{align*}
This roughly coincides with the example given in Section \ref{sec:scalings}, since here the particles are embedded at their lattice sites, \ie , we use $x_j=j$.  The main differences are the missing time-average  and the different scaling of the fields.  The above choice of scaling is determined by requiring that the fluctuation fields should have nontrivial limits.

The fields $\mathcal{R}_t^N$ and $\mathcal{Y}_t^N$ can also be interpreted as time-dependent distributions. We argue in \cite{bkll11b} that $\mathcal{R}_t^N\to \mathcal{R}_t$ and $\mathcal{Y}_t^N\to \mathcal{Y}_t$, in the sense of distributions as $N\to \infty$, 
and that the limit distributions solve the following stochastic differential equations:
\begin{align}
%\begin{cases}
&\partial_t \mathcal{R} = \gamma^{-1} \partial_x^2\,  \mathcal{R} -  \partial_x \left( c W_{1}\right)\, ,\\
&\partial_t \mathcal{Y} = (2 \gamma)^{-1} \left( \partial_x^2  (u \mathcal{R}) + \partial_x^2  \mathcal{Y} \right) - 
  \partial_x \bigl( c u W_1 + c \sqrt{T/2} W_2 \bigr)\, .
%\end{cases}
\end{align}
Here $W_{1}$, $W_2$ are independent space-time white noises, $T=T(x,t)=\vep(x,t)-\frac{1}{2}u(x,t)^2$, and $c=c(x,t)=\sqrt{2 \gamma^{-1} T(x,t)}$.
Combined together with the evolution equations for $\vep$ and $u$, these results form another predictive approximation of the evolution of energy and deformation densities in the original system, however, with a greater ``resolution'' than if the fluctuation fields are neglected.
These stochastic evolution equations are derived using similar techniques and assumptions as those for the hydrodynamic equations (\ref{eq:ueq})--(\ref{eq:ebc}).  Most notably, LTE is assumed to hold.  

Controlling the fluctuation fields makes it possible to predict also the total energy fluctuations at the NESS which can then be compared with the numerical simulation results mentioned in Section \ref{sec:energyfluc}.  
As $t\to \infty$, the solutions to the hydrodynamic equations satisfy $u(x,t)\to 0$ and $\vep(x,t) \to \overline{T}(x)$ where $\overline{T}(x)$ denotes the linear profile connecting the boundary heat bath temperatures $T_{\rm L}$ and $T_{\rm R}$.  The fluctuation fields converge into two independent Gaussian fields, $\mcss{R}$ and $\mcss{Y}$, with covariances
\begin{align*}
&\mean{\mcss{R}(F)^2} =  \int_{0}^1 \!\rmd x\,  \overline{T}(x) F^{2} (x) \, ,\\
&\mean{\mcss{Y}(G)^2} = \int_{0}^1\! \rmd x\, \overline{T}(x)^2 G(x)^2 %\\  \qquad
 + (T_{\rm L}-T_{\rm R})^2 \int_{0}^1\! \rmd x\, G (x) ((-\Delta_0)^{-1} G) (x) \, ,
\end{align*}
where $\Delta_0$ denotes the Laplacian with Dirichlet boundary conditions on $[0,1]$.  (Somewhat surprisingly, an analogous computation with \defem{pinning} yields the \defem{same} NESS energy fluctuation field $\mcss{Y}$.)  
We then find the following explicit prediction for $s_\infty$, the thermodynamic limit of $s_N$ evaluated at the NESS,
\begin{equation}
s_\infty = \frac{4 T_{\rm L} T_{\rm R} + \frac{5}{3} (T_{\rm L}-T_{\rm R})^2}{(T_{\rm L}+T_{\rm R})^2} \, .
\end{equation}
Inserting the boundary values used in the simulation thus yields $s_\infty \approx 1.403$ in perfect agreement with the numerically observed value.  This is probably our strongest evidence at the moment that the above fluctuating hydrodynamics indeed forms a predictive approximation of the hydrodynamic properties of the original lattice system.

\section*{Acknowledgements}

I would like to thank NORDITA and the organisers of the workshop for providing a lively environment for discussions.  Among others, I am particularly indebted to Herbert Spohn and Cedric Bernardin for helpful discussions.  This work has been supported by the Academy of Finland and partially by the European Science Foundation.

%\section*{References}

\end{document}